\documentclass[12pt]{article}

\usepackage[cp1251]{inputenc}
\usepackage{mathrsfs}
\usepackage{graphicx}
\usepackage{amsfonts}
\begin{document}

\setcounter{page}{1}

\vspace{2cm}

\begin{center}
{\Large \bf To the Origin of Anomalous Torque Acting
on a Rotating Magnetized Ball in Vacuum}
\end{center}

\vspace{.6cm}

\begin{center}
V.S.Beskin$^{1,2}$, A.A.Zheltoukhov$^{1}$, A.K.Obukhova$^{2}$, and E.E.Stroinov$^{2}$\\
\vspace{.5cm}
{\it 
1. Lebedev Physical Institute, Russian Academy of Sciences,
Leninskii prosp. 53, \\
Moscow, 119991 Russia\\
2. Moscow Institute of Physics and Technology,
Institutskii per. 9, \\
Dolgoprudnyi, Moscow Region, 141700 Russia}\\
\end{center}

\vspace{.5cm}

Bulletin of the Lebedev Physics Institute, 2013, Vol. 40, No. 9, pp. 265-267. \\

\vspace{2cm}

{\small
\righthyphenmin=2
{\bf Abstract} --- The anomalous torque acting on a rotating magnetized ball in vacuum is 
revisited. Its value is shown to depend on the magnetic field structure inside the body}.
\noindent

\vspace{.5cm}

\vspace{1cm}

{\bf Introduction}.

The simplest model describing neutron star magnetosphere is the vacuum model~\cite{Deutsch1955, 
OstrikerGunn1969}. According to this model a neutron star is a well-conducting magnetized ball 
rotating in vacuum. The major energy release is here due to magnetodipole radiation leading to 
rotation deceleration and to reduction of the angle $\chi$ between the rotation axis and the 
magnetic moment $\mathfrak{m}$~\cite{DavisGoldstein1970}.

In spite of the fact that the vacuum model has long been known, some points have not yet been
completely clarified. In particular, at the present time no consensus of opinion exists concerning
the so-called anomalous torque, i.e., the torque acting in the direction perpendicular to the
($\mathfrak{m}{\bf \Omega}$)-plane and causing the rotation axis precession. It is called so 
because its value
\begin{equation}
\label{K}
K = \xi \frac{\mathfrak{m}^2}{R^3}\left(\frac{\Omega R}{c}\right)^2 \sin\chi \cos\chi
\label{Ky}
\end{equation}
(with $R$ as the ball radius and $\xi$ as a numerical coefficient of the order of unity) turns out 
to be $(\Omega R/c)^{-1}$ times greater than the braking torque. Different authors give different 
$\xi$ values, namely, $\xi = 1$~\cite{DavisGoldstein1970, Goldreich}, $\xi = 1/5$~\cite{GoodNg1985}, 
and $\xi = 3/5$~\cite{Melatos2000}  (see also paper~\cite{MestelMoss2005} which however are sure 
not to allow for the contribution of the electric field). On the other hand, according 
to~\cite{Michel1991, Istomin} the anomalous torque is equal to zero ($\xi = 0$), and therefore 
such a precession must be absent. The present paper is aimed at clarification of this issue and 
to calculation of the anomalous torque acting onto a rotating magnetized ball in vacuum. 

{\bf Method of calculation of the torque}.

To determine the anomalous torque, it is necessary to find the bulk and surface currents and 
charges related to ball rotation. Below the ball is thought of as ideally conducting, so that 
the freezing-in condition be met in it:
\begin{equation}
\label{corotat}
{\bf E} + {\bf\beta}_{\rm R}\times{\bf B} = 0.
\end{equation}
Here and below ${\bf \beta}_{\rm R} = {\bf \Omega}\times {\bf r}/c$. As a result, the forces 
acting on the ball can be written as
\begin{equation}
\label{F-lor}
{\rm d}{\bf F} = \rho_{\rm e} {\bf E}~ {\rm d}V
+ [{\bf j}\times {\bf B}]/c~ {\rm d}V
+ \sigma_{\rm e} {\bf E}~ {\rm d}S
+ [{\bf I}_{\rm S} \times {\bf B}]/c~ {\rm d}S,
\end{equation}
where the first two summands correspond to the bulk contribution and the other two to the surface 
one. However, if the corotation currents ${\bf j} = \rho_{\rm e}{\bf \Omega} \times {\bf r}$ alone 
are assumed to exist in the ball volume, then, as can readily be verified, the bulk part of the 
force (3) will be zero. Then, considering that on the ball surface ${\bf r} = R\cdot {\bf n}$ 
and ${\rm d}S = R^2 {\rm d}o$, where ${\rm d}o$ is an element of the solid angle, for the total 
torque acting onto the ball surface we obtain
\begin{equation}
\label{K}
{\bf K} =\int{\bf r}\times{\rm d}{\bf F}=
\frac{c R^3}{4\pi}\int\Bigl( \left[{\bf n}\times\left\{{\bf B}\right\}\right]
\left({\bf B}\cdot{\bf n}\right)+\left[{\bf n}\times{\bf E}\right]
\left(\left\{{\bf E}\right\}\cdot{\bf n}\right)\Bigr)
{\rm d}o,
\end{equation}
where the braces indicate the field jumps on the ball surface. Thus, the problem of finding 
the torque is reduced to finding the electromagnetic field inside and outside the ball.

We shall solve the problem by the method of expansion in the parameter $(\Omega R/c)$, 
and as can be seen from relation (4) it suffices to restrict ourselves to only first-order 
terms for the electric field and second-order terms for the magnetic field. We shall use 
the well-known property of quasi-stationary configurations, when for fields depending on 
the angle $\varphi$ and the time $t$ only in the combination $\varphi - \Omega t$ the time 
derivatives can be replaced by space derivatives, as a result of which the Maxwell equations 
can be rewritten as~\cite{Book}
\begin{eqnarray}
{\bf\nabla}\times\left({\bf E}
+ {\bf\beta}_{\rm R}\times{\bf B}\right) & = & 0,
\\
{\bf\nabla}\times\left({\bf B}
-{\bf\beta}_{\rm R}\times{\bf E}\right) & = &
\frac{4\pi}{c}{\bf j}-4\pi\rho_e{\bf\beta}_{\rm R}.
\end{eqnarray}
Since both inside and outside the ball the right-hand side of the second equation appears to 
be equal to zero (inside the ball the corotation currents alone exist), we eventually have
\begin{eqnarray}
{\bf E} + {\bf\beta}_{\rm R}\times{\bf B} & = & -\nabla \psi,
\label{E}
\\
{\bf B} - {\bf\beta}_{\rm R}\times{\bf E} & = & \nabla h,
\label{B}
\end{eqnarray}
where $\psi$ and $h$ are two scalar functions which should be found from the continuity 
condition of the corresponding electric and magnetic field components and from the 
conditions ${\bf\nabla} \cdot {\bf E} = 0$ and ${\bf\nabla} \cdot {\bf B} = 0$ outside the ball.

Thus, knowing the magnetic field of order zero in the parameter $(\Omega R/c)$,
one can use equation (7) to find the electric field corresponding to the first-order 
parameter $(\Omega R/c)$. As is well-known, outside the ball it is formed of the magnetic 
dipole radiation field and the quadrupole field of charges induced in the ball. Equation (8) 
in turn allows us to unambiguously find the magnetic field of order two in the parameter
$(\Omega R/c)$. It is formed by the wave field of both magnetodipole and quadrupole radiation.

We should emphasize that the method proposed here is inapplicable to the calculation of the
moment responsible for magnetodipole radiation because it cannot distinguish between the retarded
and advanced potentials~\cite{Book}. However this uncertainty only arises at the subsequent step 
of expansion because, as we have seen, the anomalous torque (1) is $(\Omega R/c)^{-1}$ 
times larger than the braking torque directed opposite to the rotation axis. Hence, the 
procedure described above turns out to be adequate to the formulated problem.

{\bf Results}.

We shall first of all consider the case when in the zero-order in the parameter $(\Omega R/c)$
outside and inside the ball up to the inner radius $R_{\rm in} < R$ the magnetic field is a 
dipole one
\begin{equation}
\label{2.1}
{\bf B}=\frac{{3(\bf\mathfrak{m}\cdot r)}{\bf r} - {\bf\mathfrak{m}} r^2}{r^5}.
\end{equation}
Inside the inner sphere, i.e. for $r < R_{\rm in}$, we suppose homogeneous magnetic field
\begin{equation}
{\bf B} =  \frac{2{\bf \mathfrak{m}}}{R_{\rm in}^3}.
\end{equation}
In this case the anamalous torque will be fully determined by the stress acting on the inner
sphere surface $r = R_{\rm in}$. As a result, we obtain after elementary, although fairly 
labor-consuming calculations
\begin{equation}
\xi =  \frac{8}{15} -\frac{1}{5}\frac{R}{R_{\rm in}}.
\end{equation}

In the case of a uniformly magnetized ball, the magnetic field of order zero in the 
parameter $(\Omega R/c)$ inside the ball is again determined in terms of magnetic moment 
as \mbox{${\bf B} = 2 {\bf \mathfrak{m}}/R^3$}, and outside the ball it is a dipole field 
(\ref{2.1}). This means that zero-order electric currents (and, therefore, the jump of the 
magnetic field) must exist on the ball surface. Hence, in this case along with the 
first-order electric field it is also necessary to determine the second-order magnetic 
fields. As a result, we obtain
\begin{equation}
\xi = \frac{1}{3}.
\end{equation}

Thus, one can see that the anomalous torque acting on a rotating magnetized ball in vacuum is
nonzero and depends on the structure of its internal magnetic field. The difference from the 
previous calculations is obviously due to the fact that the latter ignored the internal angular
momentum of the electromagnetic field inside the star.

The authors are grateful to Ya.N. Istomin and A.A. Philippov for fruitful discussion. The work
was supported by Federal Target Program of the Ministry of Education and Science (contract no.
14.A18.21.0790).

\end{document}